\documentclass[sigconf]{acmart}
 
\AtBeginDocument{%
  }

\usepackage{CJK}
\usepackage{stfloats}
\usepackage{float}

\copyrightyear{2025}
\acmYear{2025}
\setcopyright{rightsretained}
\acmDOI{10.1145/3706599.3719831}

\acmConference[CHI EA '25]{Extended Abstracts of the CHI Conference on Human Factors in Computing Systems}{April 26-May 1, 2025}{Yokohama, Japan}
\acmBooktitle{Extended Abstracts of the CHI Conference on Human Factors in Computing Systems (CHI EA '25), April 26-May 1, 2025, Yokohama, Japan}

\acmISBN{979-8-4007-1395-8/2025/04}

\begin{document}

\title[A Case Study of Final Fantasy XIV Players' Intimate Partner-Seeking Posts]{Honey Trap or Romantic Utopia: A Case Study of Final Fantasy XIV Players' PII Disclosure in Intimate Partner-Seeking Posts}

\author{Yihao Zhou}
\orcid{0000-0002-2947-6243}
\affiliation{%
  \institution{Pennsylvania State University}
  \city{University Park}
  \state{Pennsylvania}
  \country{USA}
}
\email{yihaozhou@psu.edu}

\author{Tanusree Sharma}
\orcid{0000-0003-1523-163X}
\affiliation{%
  \institution{Pennsylvania State University}
  \city{University Park}
  \state{Pennsylvania}
  \country{USA}
}
\email{tanusree.sharma@psu.edu}
\authornote{corresponding author}

\renewcommand{\shortauthors}{Zhou et al.}

\begin{abstract}
Massively multiplayer online games (MMOGs) can foster social interaction and relationship formation, but they pose specific privacy and safety challenges, especially in the context of mediating intimate interpersonal connections. To explore the potential risks, we conducted a case study on Final Fantasy XIV (FFXIV) players' intimate partner-seeking posts on social media. We analyzed 1,288 posts from a public Weibo account using Latent Dirichlet Allocation (LDA) topic modeling and thematic analysis. Our findings reveal that players disclose sensitive personal information and share vulnerabilities to establish trust but face difficulties in managing identity and privacy across multiple platforms. We also found that players' expectations regarding intimate partner are diversified, and mismatch of expectations may leads to issues like privacy leakage or emotional exploitation. Based on our findings, we propose design implications for reducing privacy and safety risks and fostering healthier social interactions in virtual worlds.
\end{abstract}




\maketitle
\balance

\section{Introduction}
Massively multiplayer online games (MMOGs) have long served as vibrant social spaces for people to interact, form relationships~\cite{perry_2018_online}, and even seek intimacy~\cite{freeman_2016_revisiting, van_2020_intimate, Huynh_2013_magic}. From massively multiplayer online role-playing games (MMORPGs) to competitive online battle arenas~\cite{kou_2014_playing, zhang_2024_casual}, MMOGs have become sites for community building, social interaction, and relationship development. Rich social dynamics that blur the lines between playful experience and meaningful interpersonal connection~\cite{falling_2020_maloney} are fostered through diverse gameplay.

Nowadays, the intertwining nature of games and social media introduces an important yet understudied dimension to the study of interpersonal relationships in digital spaces. Live streaming platforms like Twitch and YouTube, social media platforms like Twitter, Reddit, and Weibo~\footnote{Weibo: A Chinese social platform, where users post, share, and comment on various topics. \url{https://weibo.com/} [Last Accessed: Feb 28, 2025]}, and voice-based social platforms like Discord are becoming increasingly integral to the MMOG community, enabling players to share experiences~\cite{perry_2018_online, hamilton_2014_streaming}, coordinate events~\cite{gifford_2011_starcraft, larsen_2024_destiny2}, and explore romantic relationships~\cite{steinkuehler_2006_where, van_2020_intimate, zhou_2024_collective}.

Although HCI researchers have long acknowledged the potential and advantages of online games to foster presence, communication, and collaboration in building close relationships~\cite{freeman_2016_revisiting, tong_2021_animal, freeman_2015_simulating, van_2020_intimate, boellstorff_2015_coming, nicolas_2006_alone, Huynh_2013_magic}, many of the previous works have focused on the benefits, leaving questions about the risks and dark undercurrents insufficiently addressed~\cite{frommel_2024_building}. Safety and privacy threats, such as the unintended exposure of real-world identities, and emotional exploitation, such as manipulation through role-play dynamics, are implicit and unexplored risks~\cite{Zytko_2023_dating, zhou_2024_collective} that have recently been recognized. Even more concerning is the possibility of Intimate Partner Violence (IPV) and toxic behaviors that originate from intimate interpersonal interactions in virtual worlds~\cite{maheu_2001_infidelity, frommel_2023_capital, freeman_2022_disturbing, Schulenberg_2023_birdcage}.

Exploring game-mediated intimacy is not a standalone academic exercise. Players themselves are direct stakeholders, as their emotional well-being and privacy are often on the line~\cite{frommel_2023_capital}. MMOG developers and designers of social media platforms also have a continuously growing interest in creating safer, more supportive environments that mitigate risks while enhancing user experiences~\cite{Mandryk_2023_combating, kou_2014_playing, kou_2023_Roblox, zhang_2024_harmful}. Moreover, by targeting "digital" intimacy, we aim to speak to broader questions in CSCW and HCI about the interplay between technology and interpersonal relationships in increasingly interconnected digital worlds. 

This paper focuses on Final Fantasy XIV (FFXIV), a famous MMORPG recognized and awarded for its outstanding community support and ongoing development~\cite{awards_2024_FFXIV}, with a worldwide player base of more than 66 million~\cite{mmo_population_2024_FFXIV}, as a research site to examine how players incorporate various digital platforms to seek intimate partners. Specifically, we aim to answer the following research questions:

\textbf{RQ1:} How do FFXIV players use digital platforms for intimate partner-seeking ?

\textbf{RQ2:} What risks do intimate partner-seeking behaviors have regarding privacy and safety?

The contribution of our study to the HCI and CSCW community is threefold. First, we identified the strategic use of different digital platforms by MMOG players in the formation of intimate relationships. Second, we recognized several privacy and safety risks related to game-mediated intimacy formation, such as sensitive information disclosure and privacy leakage, and examined how these risks manifest through cross-platform communication. Finally, we proposed design implications for building safer, more inclusive virtual spaces that can support intimate connections.

\section{Background}
\begin{figure}[htbp]
    \centering
    \includegraphics[width=0.32\textwidth]{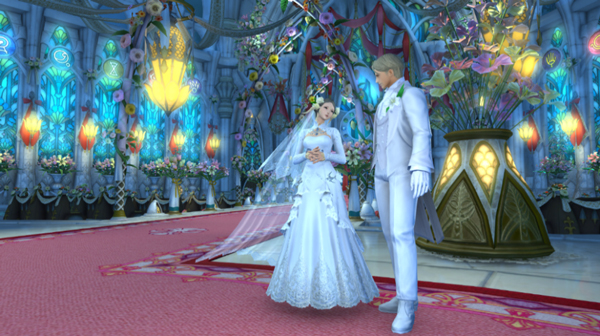}
    \caption{The Sanctum of the Twelve in FFXIV.}
    \Description{The Sanctum of the Twelve in FFXIV.}
    \label{12gods}
\end{figure}

As a renowned work in the Final Fantasy series, FFXIV has attracted millions of players worldwide since the launch of its 2013 reboot, A Realm Reborn. With its continuous content updates, by the end of 2024, FFXIV's registered player base has surpassed 66 million~\cite{mmo_population_2024_FFXIV}. Beyond its rich narrative and challenging combat, FFXIV fosters a complex social ecosystem, including Free Companies (guilds), housing districts, raid groups, and seasonal events. One notable feature is the in-game wedding system held at the Sanctum of the Twelve (see Figure \ref{12gods}) that grants couples special rewards, such as matching teleportation rings, symbolizing the game's emphasis on personal and interpersonal narratives.

\begin{figure}[htbp]
    \centering
    \includegraphics[width=0.43\textwidth]{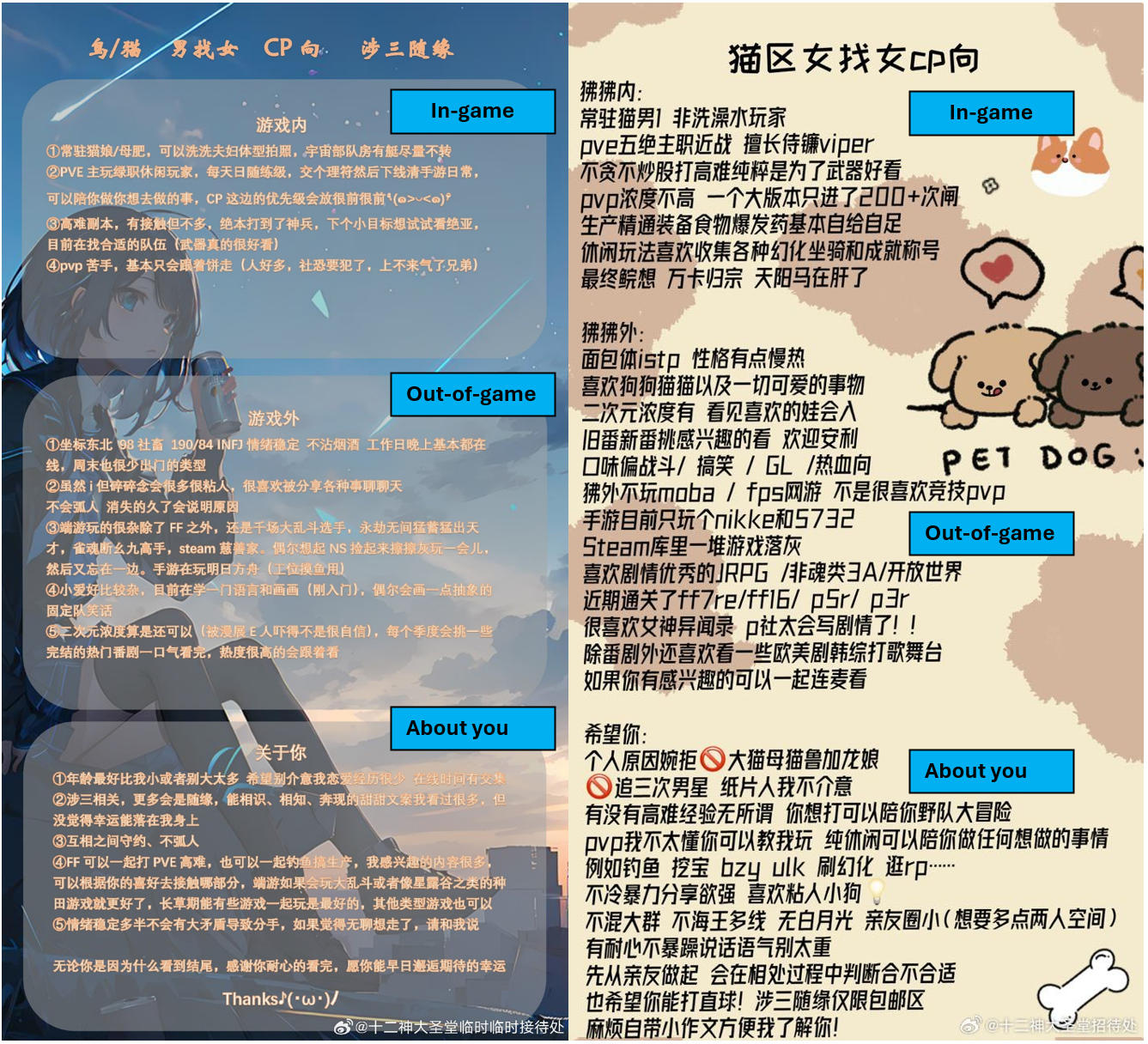}
    \caption{Examples of Recruitment Posts on a Sanctum Reception Account.}
    \Description{Examples of Recruitment Posts on a Sanctum Reception Account.}
    \label{cpdd}
\end{figure}

In the FFXIV community in China, players further extend in-game social bonds to external platforms. For example, on Weibo, a series of public anonymous accounts named "Reception Desk at the Sanctum of the Twelve" (hereafter referred to as Sanctum Reception) has emerged as a hub for intimate-partner seeking, collectively hosting over 50,000 player-submitted posts since August 2018. These posts typically include the player's data center/server, sexual orientation, interaction style, and relationship goals, often accompanied by decorated text, screenshots, and stylized cartoon images (see Figure \ref{cpdd}). The body of the posts often consists of: (i) in-game details (e.g., character body types, gameplay styles), (ii) real-world information (e.g., age, gender, personal interests), and (iii) partner expectations (both virtual and potentially offline). For example, in Figure \ref{cpdd}, the title on the right is "[Cat Server + Female seek for female + CP (couple) oriented]," followed by a descriptive narrative of the player's gaming habits, personal values, and openness to extending the relationship into the real world.

The large volume and diversity of these recruitment posts provide rich empirical grounds for examining how FFXIV players explore intimacy across interconnected digital platforms. The anonymous nature of the posts encourages candid expressions of desire, expectation, and vulnerability, which allow researchers to observe how authenticity, trust, and privacy are negotiated. Therefore, the Sanctum Reception accounts offer a valuable site for investigating how the interplay of in-game and cross-platform contexts can contribute to our broader understanding of digital intimacy and its complexities.

\section{Related Work}
\subsection{MMOGs for Socialization and Relationship Formation}

Early HCI and CSCW research has provided evidence that online games can provide similar opportunities for socializing as offline leisure activities. Steinkuehler~\cite{steinkuehler_2006_where} and Ducheneaut et al.~\cite{ducheneaut_virtual_2007} positioned MMOGs as "third places" that foster sociability through anonymity and shared context, and leading to guilds, clans, and persistent communities. Beyond community formation, Boellstorff's ethnography work of Second Life~\cite{boellstorff_2015_coming} and Huynh et al.'s work on MapleStory~\cite{Huynh_2013_magic} demonstrated how in-game cooperation and group raids can help build trust and emotional closeness, while Freeman et al.~\cite{freeman_2016_revisiting} demonstrated how dyadic gameplay in Audition can construct intimacy. These works strengthened the argument that the narrative frameworks and persistent worlds of MMOGs can create what Nardi~\cite{nardi_2006_strangers} called "collectives of strangers" who, through gameplay, become confidants, friends, and sometimes partners.

\subsection{Player Interactions on Third-Party Communication Platforms}

While virtual worlds can cultivate intimacy, the increasing interconnectivity of digital platforms makes it difficult to consider these relationships in an isolated "magic circle"~\cite{larsen_2024_Masquerade}. Scholars have shown how MMOG social and ludic elements intertwine with Twitch~\cite{Pellicone_2017_streaming, Sheng_2020_twitch}, Discord~\cite{MMO_Flor_2023, Lindsey_2022_MMORPG, poor_2019_building}, Reddit, YouTube~\cite{larsen_2024_destiny2}, and other platforms. For example, in FFXIV, players often coordinate raids, plan in-game weddings, and extend personal discussions via Discord~\cite{mortimore2020literacy, van_2020_intimate, MMO_Flor_2023, korkeila_2018_FFXIV}, while Twitch streams foster deeper viewer investment in streamers' lives~\cite{kaytoue_2012_streaming}.

The diverse affordance in multi-platform ecosystem renders intimacy-building dynamic and iterative. Ommen's ethnography study on FFXIV players revealed that digital infrastructures are essential for nurturing closeness and trust~\cite{van_2020_intimate}, and Larsen et al.~\cite{larsen_2024_destiny2} emphasize the interplay between game and online communities in strengthening relational bonds. These findings suggest that investigating game-mediated intimacy requires stepping beyond the confines of the game itself and motivate our work.

\subsection{Privacy and Safety Risks in Game-Based Intimate Relationships}

The pursuit of intimacy in digital gaming environments poses unique privacy and safety risks, particularly when relationships extend across multiple platforms~\cite{tally_2021_protect}. Social media accounts linked to game profiles may include sensitive data (e.g., real names or location histories) that were never intended for wide dissemination. This information may become a trigger for harassment, surveillance, and blackmail~\cite{zhou_2024_collective}. Furthermore, the creative freedom of role-play can give rise to emotional manipulation, deception, or even intimate partner violence (IPV) that, while distinct from offline contexts, still inflicts considerable psychological harm~\cite{Zytko_2023_dating, Ballard_2017_cyberbullying}.

Addressing these risks requires thoughtful platform governance. Designers and community managers need to balance fostering intimacy with safeguarding users from doxxing, stalking, or toxic behaviors~\cite{li_2023_LGBT, Beres_2021_toxic, freeman_2022_disturbing}. Recent efforts in academia have investigated how community moderation, codes of conduct, and robust reporting tools can shape safer interpersonal environments~\cite{Sykownik_2022_something, Grace_2022_policy}. Mandryk et al.~\cite{Mandryk_2023_combating} advocate for integrating toxicity detection and automated interventions, while Kou et al.~\cite{kou_2024_moderation, kou_2023_Roblox, kou_2024_community, kou_2024_metaverse, zhang_2024_harmful} demonstrate the role of community-driven approaches in mitigating harmful behaviors within MMOG ecosystems.

While existing measures are inspiring, they remain nascent and often reactive. Scholars argue for proactive, culturally sensitive platform design~\cite{Kaushik_2024_ad}, moderation that accommodates cross-platform fluidity, and user training to enhance privacy and well-being~\cite{Kim_2024_envision}. Balancing creative freedom, role-play authenticity, and safeguards against harassment and abuse remains a persistent challenge. Our case study offers a unique, player-centric perspective on bridging this gap.

\section{Methodology}
\subsection{Data Collection}

We employed a Weibo scraping tool~\footnote{\url{https://github.com/dataabc/weibo-crawler} [Last Accessed: Feb 28, 2025]} to collect 2,502 posts from an active Sanctum Reception account between June 17 and November 17, 2024. Each entry contained metadata (e.g., creation time) and full text. We excluded posts under 30 Chinese characters (lacking informative content) and image-only posts (due to OCR limitations). Manual checks indicated no substantive differences between text- and image-based content. After filtering, the final dataset consisted of 1,288 text-based posts, which is sufficient for content analysis and topic modeling in HCI~\cite{gao_2022_WFH, Tahaei_2020_stackoverflow}.

\subsection{Data Analysis}
As a preliminary step, we conducted a word frequency analysis to familiarize ourselves with the data. The FFXIV community is rich in specialized terms, many of which are longer than two Chinese characters (e.g., "Eorzea," the name of the fictional continent in FFXIV). To account for this, we separately counted the frequency of two-character words and words longer than two characters. After removing irrelevant conversational words (e.g., "evening," "see") that did not contribute meaningful thematic insight, we derived a terminology list.

To enhance the rigor of our study, we initially attempted a bilingual approach in Chinese and English. After translating our terminology list into English, we used the Gemma-2-9B-Chinese-Chat language model~\footnote{\url{https://huggingface.co/shenzhi-wang/Gemma-2-9B-Chinese-Chat} [Last Accessed: Feb 28, 2025]} to translate the posts. However, the frequent use of homophones, slang, and metaphors caused significant loss of nuance. Manual review confirmed that the resulting English dataset lacked sufficient quality for large-scale analysis. Consequently, we proceeded with the Chinese dataset only. Our research team includes native Mandarin speakers and fluent English speakers. The lead author is an experienced FFXIV player with six years of playing experience and deep familiarity with the gaming community, which ensured that our qualitative interpretations remained robust and well-informed.

We adopted a combined approach - Latent Dirichlet Allocation (LDA) topic modeling and qualitative thematic analysis - to analyze our data. We began by applying LDA to obtain a high-level view of the major themes in our dataset. Before performing LDA, we used Python packages (jieba for Chinese, NLTK for English) to remove stop words and segment the text into phrases and words. To improve segmentation accuracy, we integrated our custom terminology list into jieba's dictionary. Based on empirical observation, we set the number of topics to 10, as additional topics beyond this threshold became increasingly fragmented and less meaningful.

To gain richer and more contextual insights, we then conducted a qualitative thematic analysis on a random sample of 200 posts (approximately 15\%). Guided by the LDA results, we reviewed the topics, examining the words associated with each topic and the posts that exemplified them. Using the LDA-derived categories as a loose framework, we performed open coding, identifying recurring patterns, themes, and behaviors described within the posts. Two Mandarin-speaking coders separately coded 10\% of the sampled posts and then resolved differences in codes through meetings, comparisons, and discussions to reach a consensus. The lead author then completed the remaining coding, grouped lower-level codes into sub-themes, and further extracted the main themes. Finally, we organized the codes into higher-level categories.

\section{Findings}

In this section, we examine how FFXIV players utilize online platforms to seek intimate partners and the associated privacy and safety risks. Our preliminary analysis of topic modeling (Table \ref{topic-modeling}) identifies three key thematic categories: \textbf{information disclosure, relationship expectations}, and \textbf{cross-platform communication}.

\subsection{Information and Vulnerability Disclosure}
One of the main themes in the posts is information and vulnerability disclosure. Typical keywords frequently came up in this theme include BMI (body mass index) (218/1288), Infatuation (21/1288), Ex-partner (65/1288), and 4i (26/1288). 4i is an Internet slang that refers to "fourth love," a relationship where one participant is biologically male, socially female, and attracted to women, while the other is biologically female, socially male, and attracted to men. The romantic relationship between the two is called fourth love.

\subsubsection{Physical and Emotional Transparency}
Discussions about BMI (body mass index) and ex-partners point to a readiness among users to present themselves as transparent, with the goal of finding a new potential intimate partner. For instance, P(1313) provided her possible future change in residence alongside disclosure of physical condition and emotional history -\textit{`` female; INFJ/ENFJ personality type; normal voice and appearance; BMI 18.5; located in Jiangsu Province; may move towards Shanghai in the future.''} 

Some player like P(1445) disclosed real-world location -\textit{``Real-life: Based in Nanjing, Jiangning.''} Some player like P(918) showed intension for real world relationship -\textit{``Interested in a real-world relationship. Ideally, the IP address should be in Henan Province or nearby regions, with Zhengzhou/Luoyang being the best option.''} These details provide not only a view of the players' real-life plan but also convey a sense of stability or future intentions, which may attract like-minded partners. 

\subsubsection{Unconventional Identities and Practices}
Our analysis highlights the keyword "infatuation," which could reflect a deeply held interest or obsession linked with online identity, potentially serving as both a bonding mechanism and a source of conflict. Another intriguing word, "4i," reflects a particular relational configuration that challenges traditional gender and sexual norms. P(1098) posted -\textit{``Male but dress as a girl at home, voice-changing.''}

In this post, a player, who self-identified as male, revealed his practice of dressing in a feminine style at home and speaking in a female voice. This indicates how players reveal personal aspects of their identities to build trust and connections.

\subsubsection{Mental Health and Emotional Needs}

Players often share their struggles with mental well-being. Below is an example revealing a player's ongoing struggles with a mental disorder. P(1423) posted -\textit{``Suffer from depression for many years; on medication; can be emotional but can behave normally; may need some comfort, but can mostly handle it myself.''}
Others include a stark candor about their unstable conditions, as illustrated by the following post P(1075) -\textit{``Vaccine Shot: Has a mental disorder and self-harm tendencies but doesn't lash out aggressively at close ones. Currently on continuous medication orz.''}
Through these disclosures, players not only inform potential partners about their mental vulnerability but also set expectations about emotional labor and mutual care in the relationship.

\subsection{Diversified Relationship Expectations and Emotional Exploitation}
Since the goal of posts is seeking intimate partners, the expectations towards relationships are another commonly described theme. Similar to traditional online dating, there are keywords like "Feelings", "Love", and "Similarity" in the topics. To better resonate with our RQ2, we focused on keywords related to risks or negative atmospheres in this theme. Such keywords include Cold Violence (243/1288), Digital Pet (58/1288), and Entanglement (30/1288).

\subsubsection{Expectations of Intimacy and Commitment}
Mentions of "cold violence" in posts signify tensions among players when a partner fails to meet their needs and oscillates between closeness and detachment in response to interpersonal stressors. As P(348) states -\textit{``The definition of a CP (couple) is a digital pet for emotional companionship - someone you treat sincerely but have no intention of developing into a real-life intimate relationship.''}
References to "digital pet" suggest how participants conceive emotional care and attention as resources to be nurtured or neglected. The metaphor also reveals an interesting phenomenon - some players are treating their partners as "pets" rather than as living humans. 

However, the definition of "digital pet" is not unified among all posts, as another player P(1001) noted -\textit{``Although it's an electronic pet, I absolutely don't accept one online and one offline... only accept 1v1.''}

Examples here reflect the players' desire for a bounded relationship with clear commitments, which is typically included in traditional romantic relationships. The keyword "entanglement" further explains the dark side of digital intimacy, where relationship expectations are not always met, and power imbalances can lead to exploitation or manipulation.

\subsubsection{Strategic Boundary Setting}
To avoid potential exploitations, some players provide a clear definition of the relationship in their posts. For example, P(62) posted -\textit{``The definition of a CP (couple) here is not limited to a 1v1 exclusive relationship within the game. It's about being each other's top priority, above regular friends, and starting as the best of friends.
Due to personal experiences, I take relationships very seriously and place great importance on spending a long time getting to know each other online and meeting in person before entering a romantic relationship.
If you have a strong goal or expectation for romance, you may skip this post. No offense intended - it's just that I feel I might not be able to meet your expectations, and this is both to avoid disappointing you and to protect myself.''}

The post by P(62) sets clear boundaries and emphasizes a thoughtful attitude towards the relationship by prioritizing mutual understanding and connection over short-term companionship or emotional labor. By defining CP as more than just a 1v1 exclusive game-based relationship, the post reflects an attempt to counter the metaphor of "digital pets." The explicit rejection of "strong romantic goals" further depicts the player's desire to manage expectations and prevent emotional exploitation. This self-protective stance not only reflects potential personal vulnerability but also showcases the challenges players face in establishing intimate relationships within social media as a dating space.

\subsection{Cross-Platform Communication and Engagement}
A large portion of keywords in the topic is related to FFXIV, indicating that the main venue for players to engage with each other is the game itself. Nevertheless, we noticed some keywords related to other platforms, which helped answer our RQ1. Typical keywords in this theme include Steam (611/1288), PVQQ (819/1288), and Open Mic (39/1288). These words indicate that players rely on out-of-game platforms, such as Steam\footnote{Steam: A digital distribution service and storefront developed by Valve. \url{https://store.steampowered.com/} [Last Accessed: Jan 22, 2025]} (a digital game distribution platform launched by Valve), QQ (a popular instant messaging app in China), and voice-based social platforms, to interact with each other and maintain connections.

This theme reveals how romantic or intimate connections span multiple digital ecosystems rather than remaining confined to a single platform. Players draw on various tools (e.g., QQ, Discord, Steam), each with unique cultural norms and features, including voice chat, streaming, and community servers, which shape collaborative gameplay and social interactions.

\subsubsection{Activity Sharing and Companionship}
The cross-platform engagement not only increases opportunities for connection and discovery but also complicates personal boundary maintenance and identity management. Players must understand the nuances of each platform in order to juggle between public and private personas to create companionship and meaningful relationships. The sharing functions on these platforms are often used to fulfill this goal. A player P(90) wrote -\textit{``You can also AFK\footnote{AFK: Abbreviation for "away from keyboard".} and play other games, or just have the mic and screen share on while resting.''} This example showcases the willingness of a player to share mic and screen time even while not playing FFXIV, demonstrating that players can maintain relationships outside the game through sharing activities and voice interactions. 

Many players specify their preferences and expectations in posts to ensure precise matchmaking. For example, P(253) wrote -\textit{``: I'm open to FPS and MOBA games, enjoy playing cooperative games on Steam, and any new releases. Regular voice chatting is essential (via KOOK\footnote{KOOK: A Chinese voice-based social platform. \url{https://www.kookapp.cn/} [Last Accessed: Jan 22, 2025]}, Oopz\footnote{Oopz: A Chinese voice-based social platform. \url{https://oopz.cn/} [Last Accessed: Jan 22, 2025]}, or Discord). I can stream games for you, or you can stream for me.''} This post shows how players express specific needs and boundaries while using multiple platforms to create tailored experiences for themselves and their companions.

\subsubsection{Public to Private Transitions}
Instant messaging applications play important roles in cross-platform engagement. Exchanging and sharing contacts or account numbers are common practices, with some posts directly including detailed account information. For instance:
\textit{``
P(519): Q: xxxxxxxxx V: xxxxxxxxxx.
P(1024): Contact: QQ xxxxxxxxxx, WeChat xxxxxxxxxxxxx. 
P(521): Steam friend number xxxxxxxxx.\footnote{To protect Personally Identifiable Information (PII), we anonymized the original post.}''}

While exposing QQ, WeChat, or Steam accounts on social media may cause unwanted harassment, players decide to include them in posts to streamline interactions, Such a practice reflects a balance between public announcements and the initiation of private, interpersonal connections.

\section{Discussion}
\subsection{Sensitive Information Disclosure}
In our findings, we observed that many players actively share their weaknesses and vulnerabilities, a behavior distinct from traditional online dating practices ~\cite{zykto_2014_impression, sharabi2021online}. Previous work has provided empirical evidence that users in traditional online dating settings tend to emphasize their strengths and conceal their weaknesses~\cite{zykto_2014_impression, sharabi2021online}. However, in our case, an opposite situation appears. 

This information-sharing behavior can be attributed to multiple factors. From a basic perspective, games provide a trusted environment, as supported by prior research~\cite{freeman_2021_hugging, li_2023_LGBT}. Such an environment fosters a sense of safety that is rarely replicated in real-world interactions. Another notable reason is the players' diverse perceptions of relationships. As our findings illustrate, some players perceive their partners as "digital pets" rather than as romantic partners. In other words, they are seeking a "situationship" that offers emotional or social support rather than a committed, long-term relationship. While at the beginning, the pseudonymity in gaming environments facilitates sensitive information disclosure~\cite{tally_2021_protect, sharma2024m} for these players, the dynamic shifts when these interactions transition to social media or when the relationship deepens.

In such cases, shared information can backfire. For instance, vulnerabilities could be treated as fodder for small talk and spread to a wider audience, and the information provider may fall victim to stigmatization~\cite{Andalibi_2020_disclosure}, psychological abuse~\cite{zykto_2014_impression}, stalking~\cite{zhou_2024_collective}, or other kinds of intimate partner violence (IPV) as a consequence of their earlier disclosures~\cite{Zytko_2023_dating}. These harmful issues become even more severe and difficult to address in terms of accountability when the situation occurs across multiple platforms.

\subsection{Multi-Platform Privacy and Identity Management}

While games often foster collaboration and intimacy, their pseudonymous nature can impede direct real-world connections. Consequently, FFXIV players frequently supplement in-game interactions with external platforms (e.g., Weibo, QQ, Steam, voice-based social platforms) that offer deeper communication but introduce more pronounced privacy concerns. Our analysis reveals how multi-layered identities and privacy management become critical as players balance openness with preserving personal boundaries. As an example, the players in our findings who publicly share their location or contact details may risk phishing, ransomware attacks, or misinformation, as argued in relevant literature.~\cite{xiao_2017_location, sharma2023user, sharma2020analysis}.

Although multichannel communication enhances spontaneity and enriches social engagement~\cite{madson2015understanding, sharabi2021online}, it can blur boundaries between public and private spheres and undermine the "appropriate flow of information" proposed in the theory of contextual integrity in privacy~\cite{nissenbaum_2004_CI}. Leaked databases, such as those for QQ, can threaten Personal Identifiable Information (PII) such as phone numbers, social networks, or even official identity documents~\cite{zhang_2019_QQ}. Thus, we argue that transitioning relationships across platforms introduces heightened vulnerability and can lead to inconsistent self-presentation or mismatched expectations - a phenomenon illustrated by the "digital pets" metaphor and players' unaligned emotional caregiving demands.

\subsection{Design Implications}
Our findings highlight the importance of addressing the risks and complexities associated with multi-layered identity and privacy management in cross-platform communication among FFXIV players. These challenges necessitate thoughtful design interventions to ensure safer, more inclusive, and engaging digital ecosystems. Hereby, we outline several design implications derived from our analysis.

\textbf{Unified Privacy Settings.}
Implementing centralized privacy controls that synchronize user preferences across platforms can streamline identity disclosure and content sharing. Granular identity management tools, such as context-specific profiles, further enable selective disclosure of personal information according to relationship type or platform.

\textbf{Trust-Building Mechanisms.}
Incorporating features such as verified profiles, mutual information exchange, and temporary communication modes (e.g., disappearing messages) can mitigate risks of deception and exploitation. These mechanisms can foster mutual trust while preserving the exploratory dynamics of cross-platform interactions.

\textbf{Privacy Education and Consent Management.}
Embedding user education and consent tools directly into platform interfaces is essential for enhancing awareness and accountability. For instance, permissions for shared content and alerts for unauthorized usage can strengthen user control. Adopting a privacy-by-design philosophy~\cite{Schaar_2010_PBDD} ensures user safety and supports the fluid yet secure connections that define cross-platform communication.

\section{Conclusion and Future Work}
In this study, we analyzed 1,288 intimate partner-seeking posts on Weibo from FFXIV players to explore the potential risks of game-mediated intimacy. Our findings provide a view of practices and challenges in forming game-mediated intimate relationships, such as sensitive information disclosure, diverse relationship expectations, and cross-platform communication, and emphasize the importance of balancing self-expression with robust privacy safeguards.

We propose design strategies (e.g., unified privacy settings, trust-building mechanisms, and integrated privacy education) to empower users and foster safer digital ecosystems. Beyond platform design, our work contributes to broader discussions into how digital tools shape relationships and intimacy. As gaming and social spaces increasingly overlap, future research can further explore cultural differences in digital intimacy and the role of moderation in mitigating risks, ultimately guiding the creation of more secure and inclusive virtual communities.


\bibliographystyle{ACM-Reference-Format}
\bibliography{main}

\begin{appendix}
\section{Topic Modeling Results}

\renewcommand{\arraystretch}{1.5}
\begin{table}[htbp]
\small
\centering
\begin{tabular}{cp{0.8\linewidth}}
\toprule
\textbf{Topic} & \textbf{Keywords} \\
\midrule
1 & First week, Chinese server, Savage, Online romance, Casual, Final Fantasy (FF), Mana\footnotemark[9], BMI, Switch, CP \\
2 & Personality, Body type, Actual identity, Onsal Hakair\footnotemark[10], Clingy, Player, Job, Imprisonment, Online, Gameplay style \\
3 & FF, Console, Series, PVQQ, Share, Work, Capability, Corporate slave, Content, High-intensity \\
4 & Umineko\footnotemark[11], Server region, No transfer, Infatuation, Age, Appearance, Later, Open Mic, Priority, 7 PM \\
5 & Shanghai, Future, Puppy, Final, Home, Goal, Silver, Seen, Weight, Daring not \\
6 & Body type, Gameplay style, High difficulty, Corporate slave, Steam, Battlefield, PVQQ, Getting along, Glamour\footnotemark[12], High-intensity \\
7 & Feelings, Love, Suitable, Serious, Direct, Ex-partner, Getting along, Response, Entanglement, Cold Violence \\
8 & Intention, Universe\footnotemark[13], Willingness, Digital pet, Providence\footnotemark[14], Process, Fixed Play, La Noscea\footnotemark[15], Forced, Whisperwind Cove\footnotemark[16] \\
9 & Magic City\footnotemark[17], Transferable, Preference, 4i, Homeowner, International, Voor Sian\footnotemark[18], Online romance, Intention, Universe \\
10 & Relationship, Confirmation, Initiative, Ability, Special, Infatuation, International, Relationship, Similarity, Straightforward \\
\bottomrule
\end{tabular}
\caption{Topic Modeling Results of the Intimate Partner-Seeking Posts}
\label{topic-modeling}
\end{table}

\footnotetext[9]{The name of a data center in FFXIV's Japanese server.}
\footnotetext[10]{A player versus player (PvP) map in FFXIV.}
\footnotetext[11]{{\begin{CJK}{UTF8}{min}ウミネコ\end{CJK} in Japanese. It refers to a species of black-tailed gull and is also the name of a data center in FFXIV's Chinese server.}}
\footnotetext[12]{A vanity system in FFXIV that allows the player to transform the appearance of their gear into that of other gear.}
\footnotetext[13]{Here is the abbreviation of the name of a data center in FFXIV's Chinese server.}
\footnotetext[14]{The name of a data center in FFXIV's Chinese server.}
\footnotetext[15]{The name of a data center in FFXIV's Chinese server.}
\footnotetext[16]{The name of a data center in FFXIV's Chinese server.}
\footnotetext[17]{A homograph. One meaning is the name of a data center in FFXIV's Chinese server. The other is the homonym of Shanghai, China.}
\footnotetext[18]{The name of a data center in FFXIV's Chinese server.}
\end{appendix}

\end{document}